\documentclass[aps,superscriptaddress,prl,twocolumn,amsmath]{revtex4-1}
\usepackage{amssymb}
\usepackage{amsmath}
\usepackage{graphicx,subfigure,float}
\usepackage[percent]{overpic}
\usepackage{dcolumn}
\usepackage{bm}
\usepackage{color}
\usepackage{array}
\usepackage{hyperref}
\usepackage[mathlines]{lineno}
\begin{document}
\title{Computational discovery of spin-polarized semimetals in spinel materials}
\author{Shenda He}
 \affiliation{Hunan Provincial Key laboratory of Thin Film Materials and Devices, School of Material Sciences and Engineering, Xiangtan University, Xiangtan 411105, China.}
 \affiliation{School of Materials Science and Engineering, Xiangtan University, Xiangtan 411105, China.}
 \author{Ruirong Kang}
 \affiliation{Hunan Provincial Key laboratory of Thin Film Materials and Devices, School of Material Sciences and Engineering, Xiangtan University, Xiangtan 411105, China.}
 \affiliation{School of Materials Science and Engineering, Xiangtan University, Xiangtan 411105, China.}
 \author{Pan Zhou}
 \email{zhoupan71234@126.com}
 \affiliation{Hunan Provincial Key laboratory of Thin Film Materials and Devices, School of Material Sciences and Engineering, Xiangtan University, Xiangtan 411105, China.}
 \author{Zehou Li}
 \affiliation{School of Materials Science and Engineering, Xiangtan University, Xiangtan 411105, China.}
 \author{Yi Yang}
 \email{yangyi@xtu.edu.cn}
 \affiliation{School of Materials Science and Engineering, Xiangtan University, Xiangtan 411105, China.}
 \author{Lizhong Sun}
 \affiliation{Hunan Provincial Key laboratory of Thin Film Materials and Devices, School of Material Sciences and Engineering, Xiangtan University, Xiangtan 411105, China.}
 \affiliation{School of Materials Science and Engineering, Xiangtan University, Xiangtan 411105, China.}
\date{\today}
\begin{abstract}
The materials with spin-polarized electronic states have attracted a huge amount of interest due to their potential applications in spintronics. Based on first-principles calculations, we study the electronic characteristics of a series of AB$_2$X$_4$ chalcogeniden spinel structures and propose two promising candidates, VZn$_2$O$_4$ and VCd$_2$S$_4$, are spin-polarized semimetal materials. Both of them have ferromagnetic ground states. Their bands near the Fermi level are completely spin-polarized and form two types of nodal rings in the spin-up channel, and the large gaps in the spin-down channel prevent the spin-flip. Further symmetry analysis reveals that the nodal rings are protected by the glide mirror or mirror symmetries. Significantly, these nodal rings connect with each other and form a nodal chain structure, which can be well described by a simple four-band tight-binding (TB) model. The two ternary chalcogeniden spinel materials with a fully spin polarized nodal chain can serve as a prominent platform in the future applications of spintronic.\\
\end{abstract}
\maketitle
\section*{\uppercase\expandafter{\romannumeral1}. INTRODUCTION}
\indent Since the discovery of topological insulators, topological states of matter have attracted a lot of interest because of their extraordinary transport, magnetic, and optical properties.\cite{konig2007quantum,lei2020magnetized,tokura2019magnetic,yu2010quantized,chang2013experimental,PhysRevB.85.165110,nagaosa2020transport,zhang2018nodal} Topological semimetals are materials that exhibit nontrivial band crossings near the Fermi level and can be classified according to the dimension and distribution of the crossing points: nodal points (zero-dimension), nodal lines and nodal rings (one-dimension), and even nodal surfaces (two-dimension)\cite{zhang2021weyl,wu2018nodal,liang2016node,zhang2018nodal,yan2017nodal,soluyanov2015type}. Interestingly, the nodal line can be twisted into a number of shapes, resulting in more complicated topological phases such as nodal chain, Hopf-link states, and nodal-net.\cite{li2020interlocking,PhysRevB.96.201305,PhysRevB.96.041102,li2019new}

\indent Topological semimetals have already been identified in a series of materials, including IrF$_4$,\cite{bzduvsek2016nodal} Ba$_3$Si$_ 4$,\cite{cai2018nodal} carbon network,\cite{li2020interlocking} and $\beta$-cristobalite BiO$_2$\cite{PhysRevLett.108.140405}. The majority of them, however, are nonmagnetic, whereas magnetic semimetals appear to exhibit exotic features such as tunable nodal points and anomalous Hall effect. Magnetic topological semimetal was initially proposed in pyrochlore iridates AIr$_2$O$_7$ (A is Y or rare-earth element)\cite{PhysRevB.83.205101}, and a ferromagnetic (FM) material(HgCr$_2$Se$_4$) was later proposed to be a novel Weyl semimetal\cite{PhysRevLett.107.186806}. Soon after, HgCr$_2$Se$_4$ has been experimentally confirmed as a half-metal by Guan et al\cite{PhysRevLett.115.087002}. Recently, more materials having magnetic topological properties have been proposed. For example, Li$_3$(FeO$_3$)$_2$ has been proposed as half-semimetal with two independent Wely-loop\cite{PhysRevB.99.075131}, and CaFeO$_3$ is theoretically predicted with line-surface electronic states\cite{zhang2021weyl}. Fully spin polarized nodal points, lines, and surfaces coexist\cite{PhysRevB.103.085135} in quasi-one-dimensional compounds XYZ$_3$ (X = Cs, Rb, Y = Cr, Cu, Z = Cl, I). These examples provide as useful starting points for further research into unique characteristics of topological states in magnetic systems.

\indent Due to its various applications, the ternary chalcogeniden spinel family is well-known and has been extensively researched for decades.\cite{li2019new} They have face center cubic structure and is represented as AB$_2$X$_4$, where A and B are metal atoms that center the X (chalcogens) tetrahedrons and octahedrons, respectively. Almost all main group and transition metal elements may be synthesized in a stable spinel form, resulting in a diverse range of elemental compositions, electron configurations, and valence states.\cite{brik2014lattice,biagioni2014systematics} As a result of these features, the spinel has a wide range of magnetic, electrical, optical, and catalytic properties.\cite{yue2021observation,lee2019efficient,huang2019electronic,granone2018effect,dong2019revealing} Until now, a series of spinel materials, including HgCr$_2$Se$_4$, VMg$_ 2$O$_4$, LiV$_2$O$_4$, FeAl$_2$O$_4$, and NiAl$_2$O$_4$,\cite{PhysRevLett.107.186806,PhysRevB.102.155116,PhysRevB.101.121113,PhysRevB.102.195124} have been theoretically predicted to have the fully spin polarized electronic states. Transition metal spinels always have various and tunable magnetic properties, which is significant for spintronic applications. Its established synthesis technology lends itself to spintronic studies as well. Therefore, exploring ideal spinel structures with fully spin-polarized electronic states is beneficial for spintronic experimental research and pontentail applications.

\indent In this work, we find two spinel structures, VZn$_2$O$_4$ and VCd$_2$S$_4$, have completely spin-polarized electronic states near the Fermi level. In their spin-up channel, they have a series of band crossings that belong to two different types of nodal rings and are protected by mirror or gliding mirror operations. We discovered that these rings form a chain-like structure. The nontrivial topological properties of the nodal chains are confirmed by the surface states. Only small gaps are opened after considering spin-orbit coupling (SOC). Hence, our work reveals a promising material platform for studying the fundamental physics of fully spin-polarized nodal-chain, which also possesses great potential for future spintronics applications.
\section*{\uppercase\expandafter{\romannumeral2}. COMPUTATIONAL METHODOLOGY}
\indent \indent The first-principles calculations were performed by using the Vienna Ab initio Simulation Package (VASP) within the framework of density functional theory (DFT).\cite{PhysRevB.59.1758,PhysRevB.54.11169,KRESSE199615} In our calculations, we used the projector augmented wave (PAW) method and the Perdew-Burke-Ernzerhof (PBE) exchange-correlation interaction within the generalized gradient approximation (GGA).\cite{PhysRevB.50.17953,PhysRevB.59.1758,PhysRevLett.77.3865,PhysRevLett.78.1396} The cutoff energy was set to 600 eV and a $15\times15\times15$ $\Gamma$-centered k-points mesh was used. The force and energy convergence criteria were set to -0.001 eV/{\AA} and 10$^{-7}$ eV, respectively. The phonon dispersion was obtained by using the PHONOPY code.\cite{phonopy} Because of the Coulomb interaction effects, the GGA+U method was applied within the rotationally invariant DFT + U approach proposed by Dudarev $et$ $al$.\cite{dudarev1998electron} and the effective U value (U$_{eff}$ = U - J) for V was set to 4 eV. The wannier90 and WannierTools packages\cite{mostofi2008wannier90,WU2017} were used to construct the maximally localized Wannier functions for the energy bands close to the Fermi level and to calculate the local band gap and surface states, respectively.
\begin{figure*}[!htbp]
\centering
  \includegraphics[trim={0.0in 0.2in 0.0in 0.0in},clip,width=0.9\linewidth]{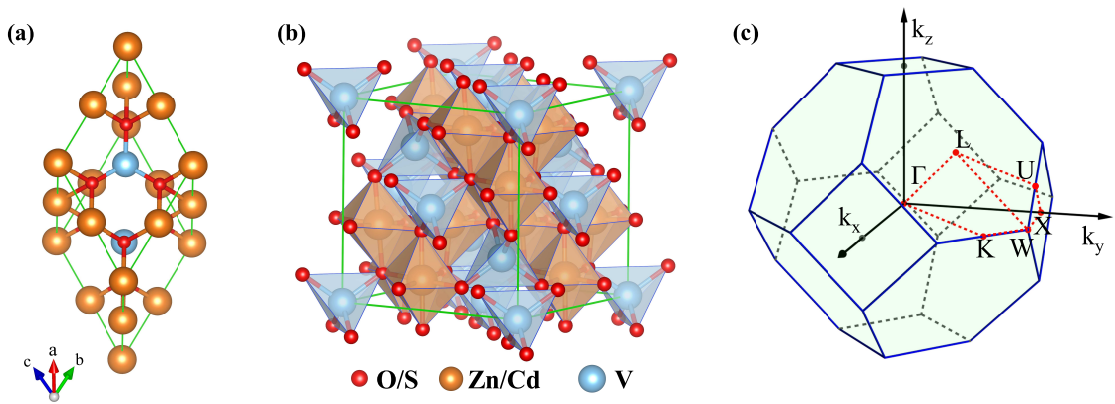}\\
  \caption{(a) Primitive cell and (b) conventional cell of VZn$_2$O$_4$ and VCd$_2$S$_4$. (c) BZ and corresponding high-symmetry paths in the first Brillouin zone.}
  \label{Fig. 1}
\end{figure*}
\section{Results and discussion}
\subsection{Material discovery}
\indent In this paper, we examine a series of spinel materials with the chemical formula AB$_2$X$_4$, where B can be Ag, Cu, Cd, Al, Ga, Ti, Ni, Ge, Hg, In, Zn or Y, A is the element of V, Mn, Fe, or Co, and X is O or S. Because most synthesized spinels have ferromagnetic (FM) ground states, we first relaxed the crystal structures and calculated their band structures in the FM state. Table S1$\dag$ summarizes their lattice constants, magnetic moments, and spin-polarized properties. According to their band structures, the most promising candidates with spin polarized semimetal states are VZn$_2$O$_4$ and VCd$_2$S$_4$. Therefore, we mainly study these two materials.
\subsection{Magnetic properties and structural stability of VZn$_2$O$_4$ and VCd$_2$S$_4$}
\indent The crystal structures of VZn$_2$O$_4$ and VCd$_2$S$_4$ are shown in Fig. 1(a) and 1(b). The V atoms are centered in the O tetrahedron, and Zn or Cd atoms are located at the centers of the O octahedron. Both of them belong to space group $Fd\bar3m$ (No. 227). The lattice constants of them are 8.534 {\AA} and 10.778 \AA, respectively. According to Bader charge analysis,\cite{HENKELMAN2006354} the O atom obtains 1.12(0.84)$\mid$e$\mid$, V atom loses 1.97(1.54)$\mid$e$\mid$, and the Zn(Cd) atom loses 1.25(0.92)$\mid$e$\mid$. Therefore, the charge transfer from V to O is greater than that of Zn(Cd). We have further confirmed the magnetic ground states of them by computing the energies of the FM and antiferromagnetic (AFM) configurations, as shown in Fig. S1$\dag$. The results show that the FM configuration is the most stable, with an energy of 70.3(86.2) meV per V atom lower than that of the AFM state in VZn$_2$O$_4$(VCd$_2$S$_4$). The magnetic moment per primitive cell in the FM ground state is 2 $\mu_B$, which is primarily contributed by V atoms with the low-spin electronic configuration(V$ \downarrow \downarrow \uparrow$). Therefore, all of the materials properties listed below are calculated from their FM ground states.

\indent To demonstrate their stability, we firstly calculate the cohesive energies (E$_c$) of VZn$_2$O$_4$ and VCd$_2$O$_4$. The results are 4.23 eV/atom and 2.86 eV/atom, respectively. In addition, we also calculate the E$_c$ of experimentally synthesized ZnV$_2$O$_4$.\cite{butt2014synthesis} The value is 3.72 eV/atom, which is comparative to that of our structures, indicating that they have the potential to be synthesized. The phonon dispersions are calculated as well, and the results are shown in Fig. 2(a) and 2(b). The absence of imaginary frequency reveals that they are kinetically stable. The first-principles molecular dynamics (MD) simulations of them are performed by using the $2\times2\times2$ supercell [see Fig. 2(c) and (d)]. The evolutions of total energies prove that the two materials are dynamically stable. Furthermore, as shown in Table S2$\dag$, the linear elastic constants and Young's modulus computed using the stresses vs. energy method meet the Born-Huang criteria (C$_{11}$-C$_{12}$$>$0, C$_{11}$+2C$_{12}$$>$0, C$_{44}$$>$0),\cite{born1955dynamical} indicating that they are mechanically stable.
\subsection{Electronic properties}
\indent Here we focus on the electronic properties of VZn$_2$O$_4$ and VCd$_2$S$_4$. Their band structures without SOC are depicted in Fig. 3. Both of their electronic states near the Fermi level are entirely spin-polarized. The bands in the spin-up channel cross the Fermi level, and show metallic properties, while the bands in the spin-down channel exhibit insulating characteristics with a large gap of 3.38 eV in VZn$_2$O$_4$ and 2.02 eV in VCd$_2$S$_4$. The bands around the Fermi level have hourglass dispersion along the high symmetry lines and form many crossing points. Their total and projected density of states are also shwon in Fig. 3(a) and 3(b), the results demonstrate that V atoms contribute the majority of their half-metallic bands near the Fermi level.\\
\indent The orbital-resolved band structures in Fig. 4(a) indicate that the V-$d_{z^2}$, V-$d_{x^2-y^2}$ orbitals contribute the most to the conduction-band minimum (CBM) and valence-band maximum (VBM) of VZn$_2$O$_4$, but there are more orbitals that contribute to the low-energy electronic states of VCd$_2$S$_4$ [see Fig. S2(a)$\dag$]. The local band gap between the CBM and VBM are calculated, as shown in Fig. 4 for VZn$_2$O$_4$ (Fig. S2$\dag$ for VCd$_2$S$_4$). It can be seen that the band crossings are not isolated, and they combine to form two different types of closed loops. We use the abbreviation NR$_1$ to represent the nodal ring that includes the crossing points on paths of $\Gamma$-K, W-$\Gamma$, and $\Gamma$-X and centers around the $\Gamma$ point in the k$_z$=0 plane. On the other hand, NR$_2$ represents the nodal ring centered around the X point in the $\Gamma$-X-L plane. Although a band crossing appears in the L-K path, it seemingly belongs to a new nodal ring in the L-$\Gamma$-K plane, which also contains the crossing in the $\Gamma$-K path. After analyzing the symmetry property of of space group $Fd\bar3m$, we discover that the L-$\Gamma$-K and $\Gamma$-X-L planes are equivalent, and the new nodal ring is equivalent to the NR$_2$.\\
\indent In order to decide the nodal rings are accidental or symmetry-protected, we further analyze the symmetries of them by calculating the irreducible representations (irreps) of the Bloch states around these crossing points (as shown in Fig. S3$\dag$). As a result, we find the NR$_1$ in the k$_z$=0 plane is under the protection of the glide mirror operation G$_z$:(x, y, z)$\rightarrow$(x + 1/4, y + 3/4, -z + 1/2), which can be verified by the corresponding opposite eigenvalues ($\pm$ 1). Similarly, the NR$_2$ in $\Gamma$-X-L plane is protected by the mirror operation G$_{\bar{1}01}$:(x, y, z) $\rightarrow$ (z, y, x) (Fig. S3$\dag$). Due to the cubic symmetry of the two materials, other equivalent rings can appear on the equivalent planes, for example, similar nodal rings NR$_1$ can appear in the planes of k$_x$ = 0, k$_y$ = 0. Surprisingly, we find that the NR$_1$ and NR$_2$ are not isolated but share the same nodal points in the paths of $\Gamma$-K and $\Gamma$-X, and form a chain-like band crossing structure in BZ in the spin-up channel, as shown in Fig. 5(a).\\
\indent Nodal chains that are topologically protected always lead to nontrivial surface states. The 2D projection of bulk BZ onto the (001) plane and the related high-symmetry points are shown in Fig. 5(a). Fig. 5(b) shows a projection of a portion of the nodal rings. We chose a path cut in the projected 2D BZ to investigate the nontrivial surface states, and the electronic local density states are shown in Fig. 5(c). The surface states show that the nodal chains are topologically nontrivial.
\begin{figure*}[!t]
\centering
  \includegraphics[trim={0.0in 0.0in 0.0in 0.0in},clip,width=6in]{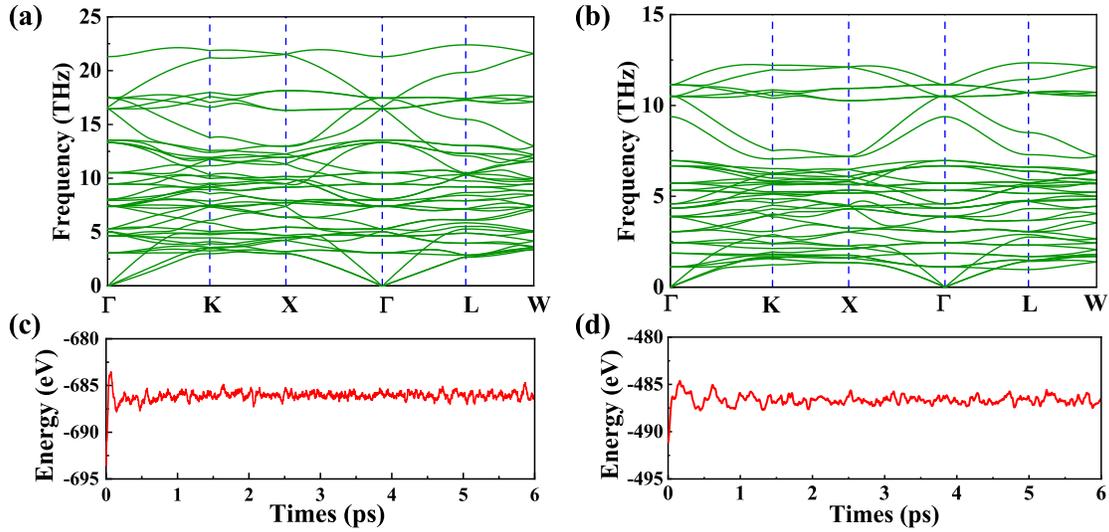}\\
  \caption{Phonon dispersions (a) for VZn$_2$O$_4$ and (b) for VCd$_2$S$_4$. The energies evolutions of them obtained by MD at 300 K for 6 ps (1 fs per step) are shown in (c) and (d), respectively. }
  \label{Fig. 2}
\end{figure*}
\begin{figure*}[!t]
\centering
  \includegraphics[trim={0.0in 0.0in 0.0in 0.0in},clip,width=0.9\linewidth]{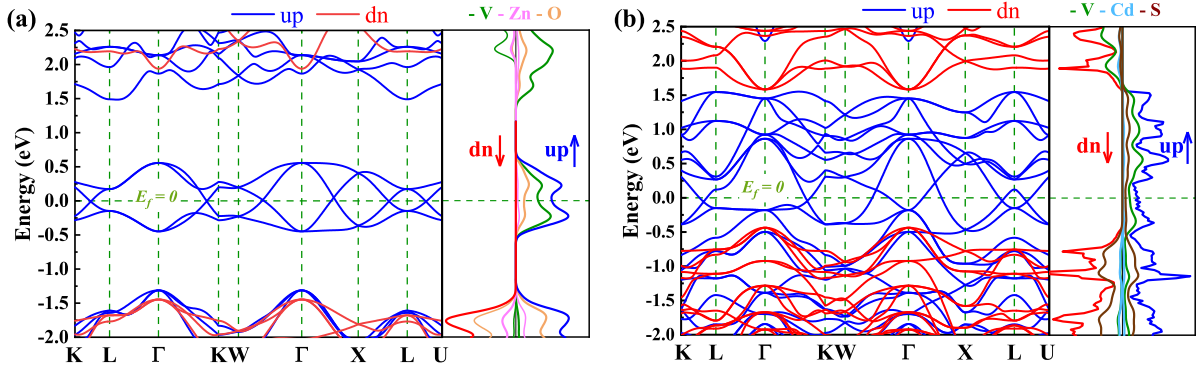}\\
  \caption{Electronic states of VZn$_2$O$_4$ and VCd$_2$S$_4$ without SOC. (a),(b) The energy band structures of them. Blue(red) lines represent spin-up(spin-down) states. The large gaps of the insulating spin-down channels are 3.38 eV and 2.02 eV, respectively. The corresponding density of states are also presented. All Fermi level are set to zero.}
  \label{Fig. 3}
\end{figure*}
\begin{figure}[!htb]
\centering
  \includegraphics[trim={0.0in 0.0in 0.0in 0.0in},clip,width=\linewidth]{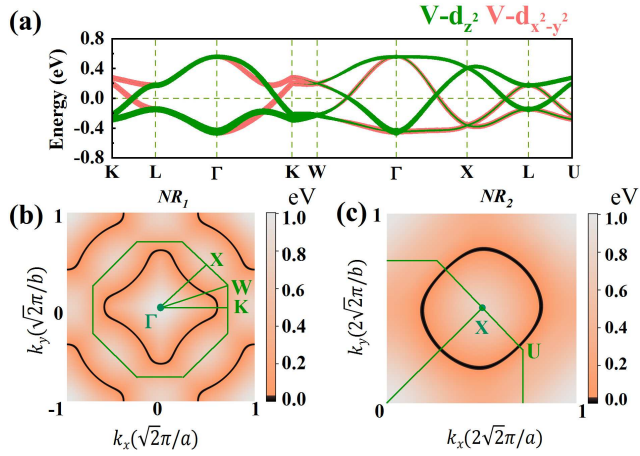}\\
  \caption{(a) Orbital-projected band structure of VZn$_2$O$_4$, including V-$d_{z^2}$ orbital (green) and V-$d_{x^2-y^2}$ orbital (orange). (b), (c) The local band gap for VZn$_2$O$_4$ between the CBM and the VBM in the k$_z$ = 0 (X-$\Gamma$-W) plane and $\Gamma$-X-U plane, respectively.}
  \label{Fig. 4}
\end{figure}
\begin{figure}[h]
\centering
  \includegraphics[trim={0.0in 0.0in 0.0in 0.0in},clip,width=\linewidth]{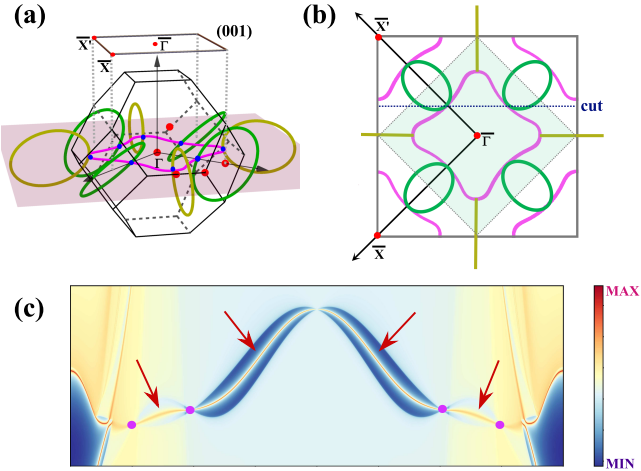}\\
  \caption{(a) Schematic illustrations of NRs in bulk BZ. The NR$_1$ is represented by purple rings, and the NR$_2$ is expressed by green or dark yellow line. The blue dots symbolise the band crossings along the high-symmetry path, which are shared by the two kinds of NR, jointly. Importantly, the NR$_1$ and NR$_2$ exist in the equivalent mirror plane. (b) The top view of 2D BZ and NRs. (c) the surface band structure along a slice path [cut in (b)], where the bulk band and surface states are represented by yellow and brown. The surface states can be seen originating from the matching band crossing (marked with purple dots).}
  \label{Fig. 5}
\end{figure}
\begin{figure}[h]
\centering
  \includegraphics[trim={0.0in 0.0in 0.0in 0.0in},clip,width=\linewidth]{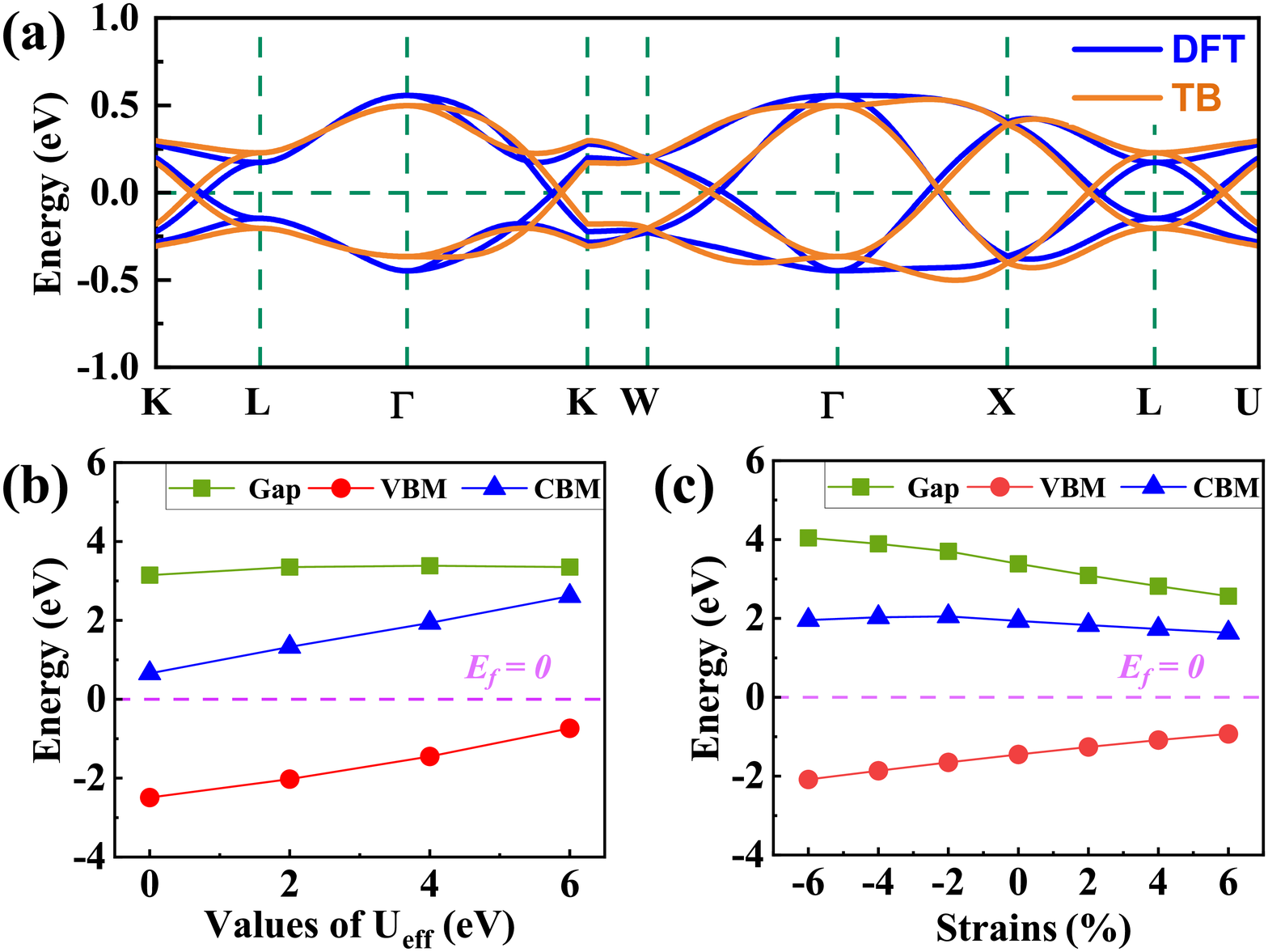}\\
  \caption{(a) Four isolated bands from the DFT calculation and TB model. (b) and (c) The values of VBM (red), CBM (blue), and gap (green) of VZn$_2$O$_4$ in the spin-down channel with U$_{eff}$ shifted or under series of hydrostatic strains.}
  \label{Fig6}
\end{figure}
\subsection{Tight-binding model}
\indent In this section, we construct a tight-binding model based on the Slater-Koster method\cite{PhysRev.94.1498} to understand the formation of the nodal chain. As seen in the orbital-projected band structures without SOC, the \emph{d$_{x^2-y^2}$} and \emph{d$_{z^2}$} orbitals of V atoms contribute the most to the four isolated band around Fermi level for VZn$_2$O$_4$. Therefore, we use them as the basis to construct a four-band tight-binding model. A Hamiltonian considering the d-d hopping until the next-nearest is given by\\
\begin{equation}
\label{eq1}
\textbf{\emph{H}}=\sum_{i}\epsilon_i+\sum_{r=1}^2(\sum_{i,j}E_{ij}^rc_{i}^{\dagger}{c}_{j}+\textrm{H}\cdot{c})
\end{equation}
\begin{equation}
\begin{aligned}
E_{1, 1}^r = &\frac{3}{4}\left(l^{2}-m^{2}\right)^{2}(d d \sigma^r)+\left[l^{2}+m^{2}-\left(l^{2}-m^{2}\right)^{2}\right](d d \pi^r)\\
&+\left[n^{2}+\frac{1}{4}\left(l^{2}-m^{2}\right)^{2}\right](d d \delta^r), \\
E_{1, 2}^r = &\frac{1}{2} \sqrt{3}\left(l^{2}-m^{2}\right)\left[n^{2}-\frac{1}{2}\left(l^{2}+m^{2}\right)\right](d d \sigma^r)\\
&+\sqrt{3} n^{2}\left(m^{2}-l^{2}\right)(d d \pi^r)+\frac{1}{4} \sqrt{3}\left(1+n^{2}\right)\left(l^{2}-m^{2}\right)(d d \delta^r), \\
E_{2, 2}^r = &\left[n^{2}-\frac{1}{2}\left(l^{2}+m^{2}\right)\right]^{2}(d d \sigma^r)+3 n^{2}\left(l^{2}+m^{2}\right)(d d \pi^r)\\
&+\frac{3}{4}\left(l^{2}+m^{2}\right)^{2}(d d \delta^r),
\end{aligned}
\end{equation}
where $\epsilon_i$ is the on-site energy at the \emph{i}th sites. \emph{r}=1 and 2 denote the nearest hopping and the next-nearest hopping interactions, respectively. c$_{i}$$^{\dagger}$(c$_j$) is the creation (annihilation) operator of electrons at site \emph{i} (\emph{j}). \emph{E$_{ij}^n$} is the \emph{n}-hopping coefficient between the \emph{i}th and \emph{j}th sites. \emph{l}, \emph{m}, and \emph{n} represent the directional cosine of \emph{R$_{ij}$} along the three directions \emph{x}, \emph{y},and \emph{z}, respectively. \emph{dd$\sigma$}, \emph{dd$\pi$}, and \emph{dd$\delta$} are independent parameters in the SK method. Because of the double degenerate energy bands along the L-$\Gamma$ path, the on-site energies of the \emph{d$_{x^2-y^2}$} and \emph{d$_{z^2}$} orbitals must be equal. Besides, after considering the concrete structure of VZn$_2$O$_4$, we find $\emph{dd$\sigma$}^1$ vanishes in the final expressions of hopping paramerters. As a result, there are six independent parameters in the TB model. After fitting the first-principles energy bands, we obtain the values of these parameters, and they are listed in tables 1. The band structures computed by the TB model Hamiltonian correspond well with those derived by the DFT, as shown in Fig. 6(a).\\
\begin{table}[h]
\caption{The Parameters of the TB Hamiltonian for VZn$_2$O$_4$}
  \label{table1}
  \renewcommand{\arraystretch}{1.7}
    \begin{tabular*}{0.47\textwidth}{@{\extracolsep{\fill}}ccc}
    \hline
    Onsite energy & $\epsilon$ & 0.012 eV \\
    \hline
    Nearest neighbor  & $dd\pi^1$ & -1.097 eV\\
                                    & $dd\delta^1$ &  1.869 eV\\
    \hline
                                             & $dd\sigma^2$ & -0.555 eV\\
    Next nearest neighbor          & $dd\pi^2$ & -0.045 eV\\
                                             & $dd\delta^2$ &  0.257 eV\\
    \hline
    \end{tabular*}
\end{table}
\subsection{Robust semimetal states and the effect of SOC}
\indent To investigate the effect of different electronic correlations on semimetal states, here, we further calculated the energy band structures and the evolution of the VBM, CBM and gaps with the U$_{eff}$ range from 0 to 6 eV, as shown in Fig. S4$\dag$ and Fig. 6(b). When the U$_{eff}$ value increases, the energy values of VBM and CBM increase, but the band gaps of spin-down bands do not change significantly. Moreover, the crossing points remain around the Fermi level with varying U$_{eff}$. It can be concluded that the semimetal characters are insensitive to electron correlation.\\
\indent Finally, we examine how the external strain and SOC affect the energy band structures of these two materials. As shown in Fig. 6(c), the band gaps of the spin-down bands become smaller with increasing lattice constants, but they are still insulated, and the semimetal states around the Fermi level are still well preserved (Fig. S5$\dag$). The band structures with SOC are presented in Fig. S6$\dag$. We find that all opened gaps are below 10 meV, which means they can always be neglected at room temperature.\\
\section{Conclusion}
\indent In conclusion, we proposed two spinel materials with nontrivial spin-polarized semimetal states: VZn$_2$O$_4$ and VCd$_2$S$_4$. Both of them have FM ground states and are kinetically and mechanically stable. The band crossings around the Fermi level belong to two kinds of fully spin polarized hourglass nodal rings and they are, respectively, protected by the glide operation G$_z$ and mirror operation G$_{\bar{1}01}$ (or their equivalent operations). The nontrivial properties are also verified by their nontrivial surface states. Remarkably, we find the two kinds of nodal rings link together and form a chain-like structure. When SOC is included, negligible gaps are opened for the semimetal states. The band crossings are also well retained under various U$_{eff}$ and external strains. These two spinel materials provide an excellent platform for studying the topological semimetal states of ferromagnetic systems.\\
\section*{Conflicts of interest}
There are no conflicts to declare.
\section*{Acknowledgements}
\indent This work is supported by the National Natural Science Foundation of China (Grant Nos. 11804287 and 11574260), the Hunan Provincial Natural Science Foundation of China (2019JJ50577, 2021JJ30686, 2019JJ60006) and Hunan Provincial Innovation Foundation For Postgraduate (CX20210623).
\bibliography{ref} 
\bibliographystyle{rsc} 
\end{document}